\documentclass[final,5p,times,twocolumn]{elsarticle}
\usepackage{lineno,hyperref}
\modulolinenumbers[5]

\journal{Nuclear Inst. and Methods in Physics Research, A}

\usepackage{xcolor}
\usepackage{soul}

\begin{document}

\begin{frontmatter}

\title{Radiation in oriented crystals: innovative application to future positron sources}



\author[ferrara,infnferrara]{Mattia Soldani\corref{corr}}
\author[orsay]{Fahad Alharthi}
\author[infnferrara]{Laura Bandiera\corref{corr2}}
\author[infnferrara]{Nicola Canale}
\author[sapienza]{Gianluca Cavoto}
\author[orsay]{Iryna Chaikovska}
\author[orsay]{Robert Chehab}
\author[ferrara,infnferrara]{Vincenzo Guidi}
\author[minsk]{Viktar Haurylavets}
\author[ferrara,infnferrara]{Andrea Mazzolari}
\author[ferrara,infnferrara]{Riccardo Negrello}
\author[infnferrara]{Gianfranco Paternò}
\author[ferrara,infnferrara]{Marco Romagnoni}
\author[infnferrara,kisti]{Alexei Sytov}
\author[minsk]{Victor Tikhomirov}

\cortext[corr]{Corresponding author; \textit{Email address:} \texttt{mattia.soldani@lnf.infn.it} (Current institution: INFN Laboratori Nazionali di Frascati, Frascati, Italy)}

\cortext[corr2]{Corresponding author; \textit{Email address:} \texttt{bandiera@fe.infn.it}}

\address[ferrara]{Università degli Studi di Ferrara, Ferrara, Italy}
\address[infnferrara]{INFN Sezione di Ferrara, Ferrara, Italy}
\address[orsay]{Université Paris-Saclay, CNRS/IN2P3, IJCLab, Orsay, France}
\address[sapienza]{Sapienza Università di Roma, Roma, Italy}
\address[minsk]{INP, Belarusian State University, Minsk, Belarus}
\address[kisti]{KISTI, Daejeon, Korea}

\begin{abstract}
It has been known since decades that the alignment of a beam of high-energy electrons with particular crystal directions involves a significant increase of bremsstrahlung radiation emission. This enhancement lies at the conceptual foundation of innovative positron source schemes for future lepton colliders. In particular, the so-called hybrid scheme makes use of a heavy-metal radiator in crystalline form, which is then followed by an amorphous metallic converter for positron generation from electrons by means of a two-step electromagnetic process. This work presents the most recent simulation results obtained on the development of a hybrid positron source for the FCC-$ee$ from the standpoint of the features of both the crystalline radiator and the amorphous converter.
\end{abstract}


\end{frontmatter}


\section{Positron source schemes}

In the development of an $e^+e^-$ collider, the generation of a positron beam at the same intensity scale as the electron one proves one of the most important and challenging tasks. Indeed, differently from electrons, positron intense beams cannot be easily obtained, and dedicated systems devoted to the positron production, collection and acceleration are required. Typically, the photon conversion of high-energy photons into $e^+e^-$ pairs in matter is exploited for applications at the current beam intensity frontier \cite{chehab_positron_1989, chaikovska_positron_2022}. In particular, the beam from the electron source (typically a linac \cite{chehab_positron_1989}) can be exploited to generate the photons that are needed for the positron production via bremsstrahlung. 

\begin{figure}[h!]
\centering
\includegraphics[width=\columnwidth]{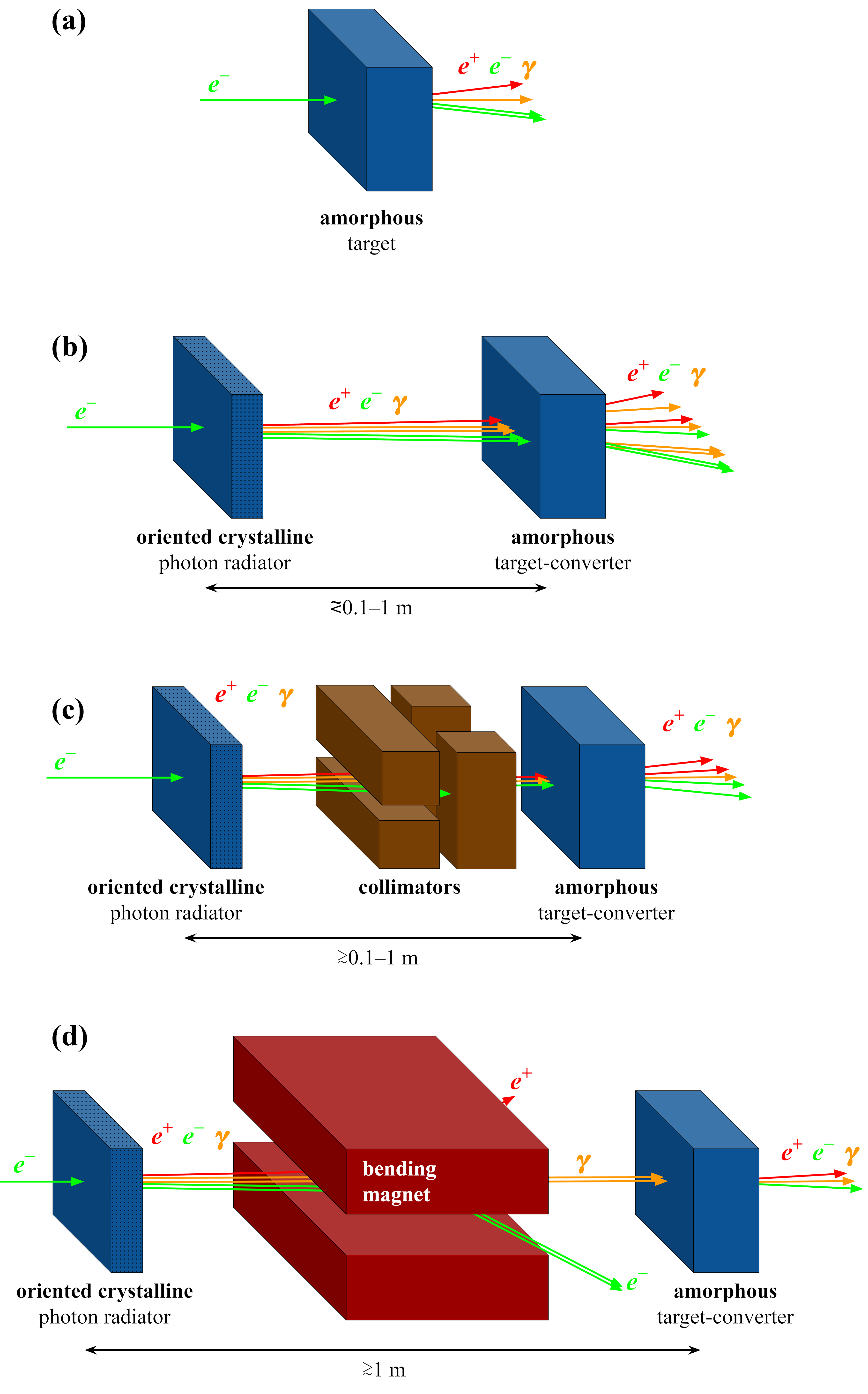}
\caption{Comparison between different positron source schemes: conventional (a), hybrid simple (b) and hybrid optimised with collimators (c) or with a bending magnet (d).}
\label{fig:ps_schemes_all}
\end{figure}

All present high-energy lepton colliders exploit the so-called conventional positron source scheme -- figure \ref{fig:ps_schemes_all}a: a fraction of the primary electrons impinges on a high-$Z$, high-density, very thick (several $X_0$) target, starting electromagnetic showers that result in the emission of photons, electrons and positrons \cite{chehab_positron_1989, bandiera_crystal-based_2022}, which can then be easily separated from one another with magnetic fields. Albeit well-known and simple, this standard scheme does not prove suitable for next-generation $e^+e^-$ colliders such as the FCC-$ee$ (Future Circular Collider, electron-electron) and the ILC (International Linear Collider). Indeed, the unprecedented intensity scale would result in serious target heating and radiation issues \cite{chaikovska_positron_2022, bandiera_crystal-based_2022, artru_polarized_2008}, especially if the target is optimised for the generation of as many positrons as possible -- $3$--$6~X_0$ in current machines at the GeV scale or higher \cite{chaikovska_positron_2022}. Moreover, the beams obtained with the conventional scheme feature an output emittance which is orders of magnitude worse than the ones from high-brilliance electron sources due to the large divergence and momentum spread resulting from the shower development and the multiple scattering in the target \cite{chaikovska_positron_2022}.

The design of future lepton colliders has driven the R\&D of novel positron source concepts that would maximise the positron production rate, minimise the output emittance and keep the energy deposit inside the bulk and its activation at reasonable levels, ultimately avoiding the target destruction.

A particularly appealing way to achieve maximum positron rate and minimum emittance with a Peak Energy Deposition Density (PEDD, i.e., the average of the event-by-event maxima of energy deposit per incident particle and per unit volume) smaller than in the conventional scheme consists in splitting the single target into two separate stages: the input electrons impinge on the upstream target (the radiator), with a thickness smaller than the radiation length, generating photons; then, the photons interact on the other, much thicker target (the converter), located at a distance downstream with respect to the radiator, where $e^+e^-$ pairs are eventually produced. This is the so-called hybrid scheme \cite{bandiera_crystal-based_2022, artru_polarized_2008, noauthor_ichep_nodate}, and is sketched in figure \ref{fig:ps_schemes_all}b. It has several advantages with respect to the single-stage positron source:

\begin{itemize}
\item the total deposited energy is shared between the two targets, therefore both receive an overall lower energy load \cite{noauthor_positron_nodate};

\item the integral energy deposit in the $\lesssim X_0$ thick radiator and the resulting PEDD are rather small \cite{noauthor_positron_nodate};

\item the integral energy deposit and the PEDD in the converter can be reduced by increasing the distance from the radiator, thus cutting the tails in the output of the latter off from the acceptance.
\end{itemize}

Moreover, collimators or magnetic fields can be placed between the two stages \cite{bandiera_crystal-based_2022}. The collimators (figure \ref{fig:ps_schemes_all}c) further contribute to cut off the tails of the upstream output beam, whereas a magnet (figure \ref{fig:ps_schemes_all}d) sweeps all the (undesired) charged particles away from the converter acceptance. In practice, all these options feature different input beam intensity tolerance, positron production rate, output emittance and radiation level in and around the two targets, etcetera -- the choice of the optimal configuration mainly depending on the accelerator performance requirements. 

As discussed in section \ref{sec:crystalline_effects}, an oriented crystalline radiator can attain the same or higher photon yield and lower output positron angular aperture than an amorphous one while being considerably thinner \cite{bandiera_crystal-based_2022}. This solution is currently under study for the FCC-$ee$: extensive measurements and simulations have been \cite{bandiera_crystal-based_2022, alharthi_target_2022} and are currently being performed on tungsten ($Z = 74$, $X_0 = 0.3504~\mathrm{cm}$ \cite{noauthor_particle_nodate}) on the features of the radiation resulting from coherent interactions by electrons of energies rather close to the value of the FCC-$ee$ primary electron beam, i.e., $6~\mathrm{GeV}$ \cite{chaikovska_positron_2022}. Tungsten oriented along the $[111]$ axis was chosen because of the extremely strong potential associated with it, i.e., $U_0 \sim 890~\mathrm{eV}$ at room temperature \cite{soldani_innovative_2023}.

\section{Coherent crystalline effects at the $\mathrm{GeV}$ scale}
\label{sec:crystalline_effects}

It has been well known since the 1950s that the electromagnetic interactions between high-energy electrons/positrons and crystals can be strongly affected by the atomic lattice structure of the latter \cite{ter_mikaelian_interference_1953}. Firstly, the so-called coherent bremsstrahlung (CB) was hypothesised \cite{ter_mikaelian_interference_1953} and then experimentally observed \cite{Palazzi68}. CB consists of the enhancement of the probability for bremsstrahlung radiation emission that occurs when the momentum transferred by the electron/positron to the crystalline bulk matches a reciprocal lattice vector, in analogy with Bragg-Laue diffraction. This enhancement is attained at rather small angles between the incident particle trajectory and a lattice symmetry (plane or axis) -- CB-related effects being observed at up to $1^\circ$ \cite{bandiera_strong_2018, soldani_next-generation_2021}.

The CB description of the interactions between electrons/positrons and crystals works as long as the nearly straight trajectory approximation can be applied to the motion of the incident particle. Conversely, this approximation proves inadequate when the incident particle trajectory is aligned with the crystal plane/axis within the so-called Lindhard critical angle \cite{lindhard65}, $\theta_\mathrm{L} = \sqrt{2 U_0 / E}$, where $E$ is the lepton energy. Under this condition, channelling occurs \cite{lindhard65, sorensen_channelling_1987, uggerhoj_interaction_2005}: the coherent interactions with the atoms in the same planes/strings force the particle into transverse oscillations in an effective electric planar/axial field of $\varepsilon \sim 10^{10}$--$10^{12}~\mbox{V}/\mbox{cm}$. Channelled particles undergo the emission of the so-called channelling radiation, i.e., the increase of different components of the electromagnetic radiation spectrum, depending on the initial energy scale: as a rule of thumb, higher initial energies result in the boost of harder spectrum components -- see, e.g., \cite{bandiera_strong_2018, bandiera_study_2018, soldani_enhanced_2022}.

In the rest frame of the incident particle, the aforementioned effective field is enhanced by a Lorentz factor, $\gamma = E / m c^2$, where $m c^2$ is the electron mass, up to the point of becoming comparable to the QED critical field, $\varepsilon_0 = m^2 c^3 / e \hbar \sim 1.32 \times 10^{16}~\mbox{V}/\mbox{cm}$ \cite{uggerhoj_interaction_2005, Sauter31} if the initial energy is sufficiently high.

This extremely intense field (namely, the strong field) is approximately constant over long sections of the particle motion inside the crystalline lattice. As a result, in the so-called constant field approximation \cite{uggerhoj_interaction_2005, baier_electromagnetic_1998, KIMBALL198425, KIMBALL198569, VGBaryshevskii_1989}, the particle propagates in the same way as under the effect of a uniform magnetic field, i.e., like in a synchrotron \cite{uggerhoj_interaction_2005}. Similarly, the radiation emission in this regime is of quantum, synchrotron type, featuring a dramatic boost of the hard part of the energy spectrum \cite{uggerhoj_interaction_2005, soldani_experimental_2022}.

The full strong field regime sets in when $\chi = \gamma \varepsilon / \varepsilon_0 > 1$ \cite{uggerhoj_interaction_2005, baier_electromagnetic_1998}, which, for instance, corresponds to a primary energy threshold of $\sim 16.3~\mathrm{GeV}$ in case of crystalline tungsten oriented along the $[111]$ axis at room temperature \cite{soldani_innovative_2023}. However, limited strong field effects are already observed at $\chi \gtrsim 0.1$ -- see, e.g., \cite{soldani_experimental_2022}. An estimate of the angular acceptance of this effect around the lattice symmetry is provided by $\Theta_0 = U_0 / m$ \cite{uggerhoj_interaction_2005, baier_electromagnetic_1998, BARYSHEVSKII1985430}, which is independent on the initial energy and several times greater than $\theta_\mathrm{L}$ at the $\mathrm{GeV}$ scale. In case of tungsten $[111]$, $\Theta_0 \sim 1.74~\mathrm{mrad}$ \cite{soldani_experimental_2022, soldani_innovative_2023}.

\section{Features of the radiator}

The simulation of the full positron source hybrid scheme is performed in two separate stages. Firstly, the interactions of the electron primary beam in the tungsten crystalline radiator are simulated with a dedicated code: the electron trajectories are calculated using the equation of motion in the crystalline potential at finite steps, and are then used in an algorithm based on the Direct Integration of the Baier-Katkov formula (DIBK) to compute the radiation emission (and also the secondary pair production). Extensive details are provided, e.g., in \cite{baryshevsky_influence_2017, guidi_radiation_2012, bandiera_radcharm_2015, doi:10.1080/10420159108220589, TIKHOMIROV1993409, PhysRevAccelBeams.22.064601}.

A crystal with a thickness of $2~\mathrm{mm}$ was chosen, as it provides a good photon yield with limited energy deposit. The $[111]$ axis was selected, as it features the strongest potential in tungsten. Details on the crystal stage optimisation can be found in \cite{bandiera_crystal-based_2022}, which shows good agreement with previous results \cite{artru_polarized_2008}.

The primary beam is a monochromatic ($6~\mathrm{GeV}$), pure electron beam, with a $500~\mu\mathrm{m}$ circular spot and a divergence of $100~\mu\mathrm{rad}$. It has been modelled after the FCC-$ee$ current design parameters. $10^4$ primary electrons have been generated, which resulted into $175672$ ($61069$) output photons (electrons/positrons). Further details on this simulation can be found in \cite{bandiera_crystal-based_2022, soldani_innovative_2023}.

\begin{figure}[h!]
\centering
\includegraphics[width=\columnwidth]{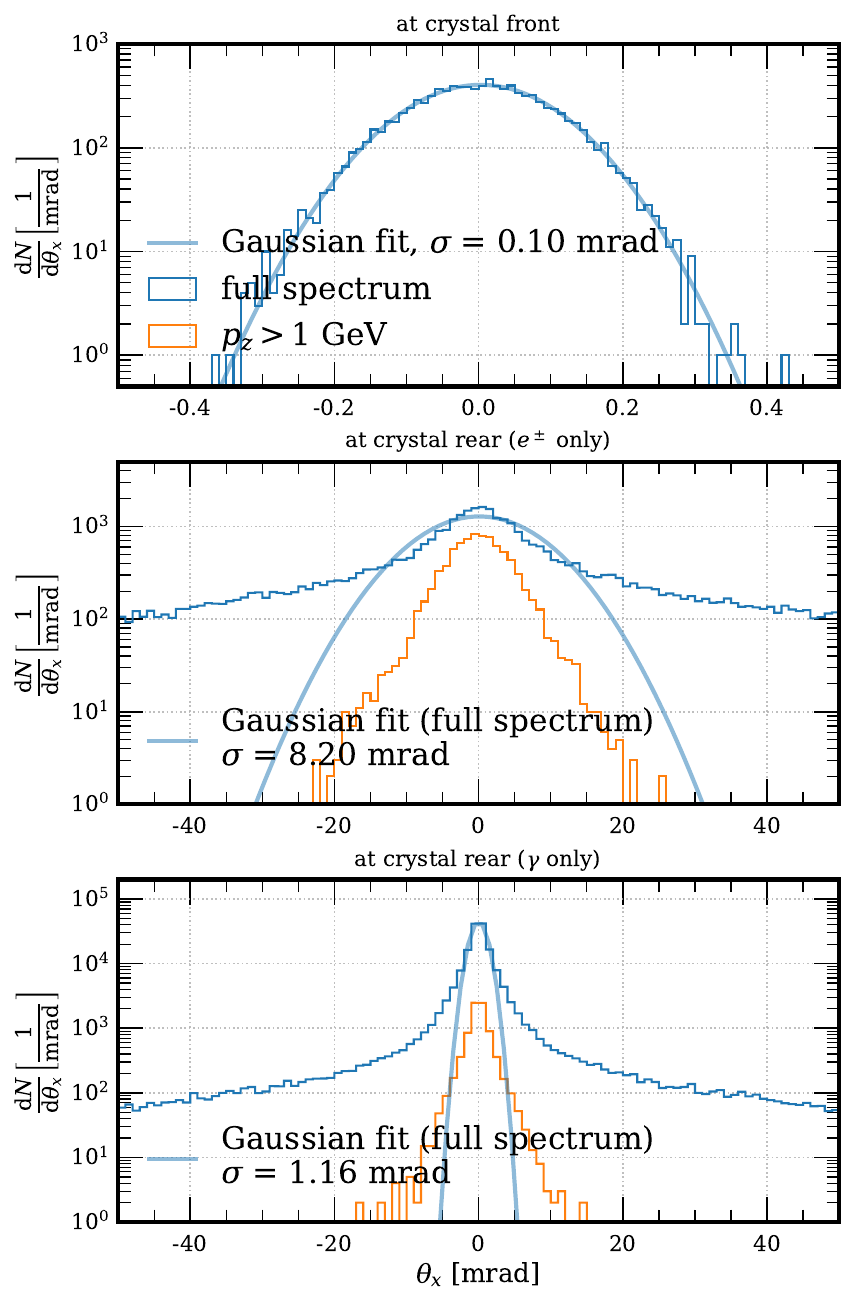}
\caption{Beam (horizontal) angular distributions at the crystal input (top) and output, for charged particles (centre) and photons (bottom). It has to be noted that the scale of the centre and bottom plots is $100$ times that of the one at the top.}
\label{fig:psFullSim_crys_angles}
\end{figure}

As expected, the beam spot size is almost entirely unaffected by the interactions inside the radiator \cite{soldani_innovative_2023}. On the other hand, the distributions of the output angles $\theta_x$ and $\theta_y$ for photons (figure \ref{fig:psFullSim_crys_angles} bottom) and for electrons/positrons (figure \ref{fig:psFullSim_crys_angles} centre) are significantly different from each other and from the input one (figure \ref{fig:psFullSim_crys_angles} top). Indeed, the divergence of the output charged component of the beam is more than $80$ times larger than the input one, mostly because of the primary electron recoil in the bremsstrahlung emission and of the multiple scattering. The large-$\theta$ tails are mostly due to small-energy electrons, as shown by the orange histogram in figure \ref{fig:psFullSim_crys_angles} centre.

On the other hand, photons emerge from the crystal with a relatively small angle with respect to the primary beam direction. The divergence increase accounts for the contributions of photons which are emitted by secondary electrons at larger angles with respect to the primary beam aperture and of the radiation cone opening angle ($1 / \gamma = 85~\mu\mathrm{rad}$). The orange curve in figure \ref{fig:psFullSim_crys_angles} bottom shows that, as for charged output particles, higher-energy photons mostly contribute to the beam core.

\section{Optimisation of the converter}

The track-by-track output of the tungsten crystal is then exploited as an input for the second simulation stage: a standard Geant4 software, in which the tungsten amorphous converter and the space (filled with vacuum) between the radiator and the converter are modelled. This stage was simulated with Geant4 version {10.7} and relies on the FTFP\_BERT reference physics list \cite{soldani_innovative_2023, noauthor_guide_nodate}. Simulations of the conventional scheme have also been performed with the same program, bypassing the crystal and directly feeding the Geant4 program with the primary electron beam tracks.

\begin{figure}[h!]
\centering
\includegraphics[width=\columnwidth]{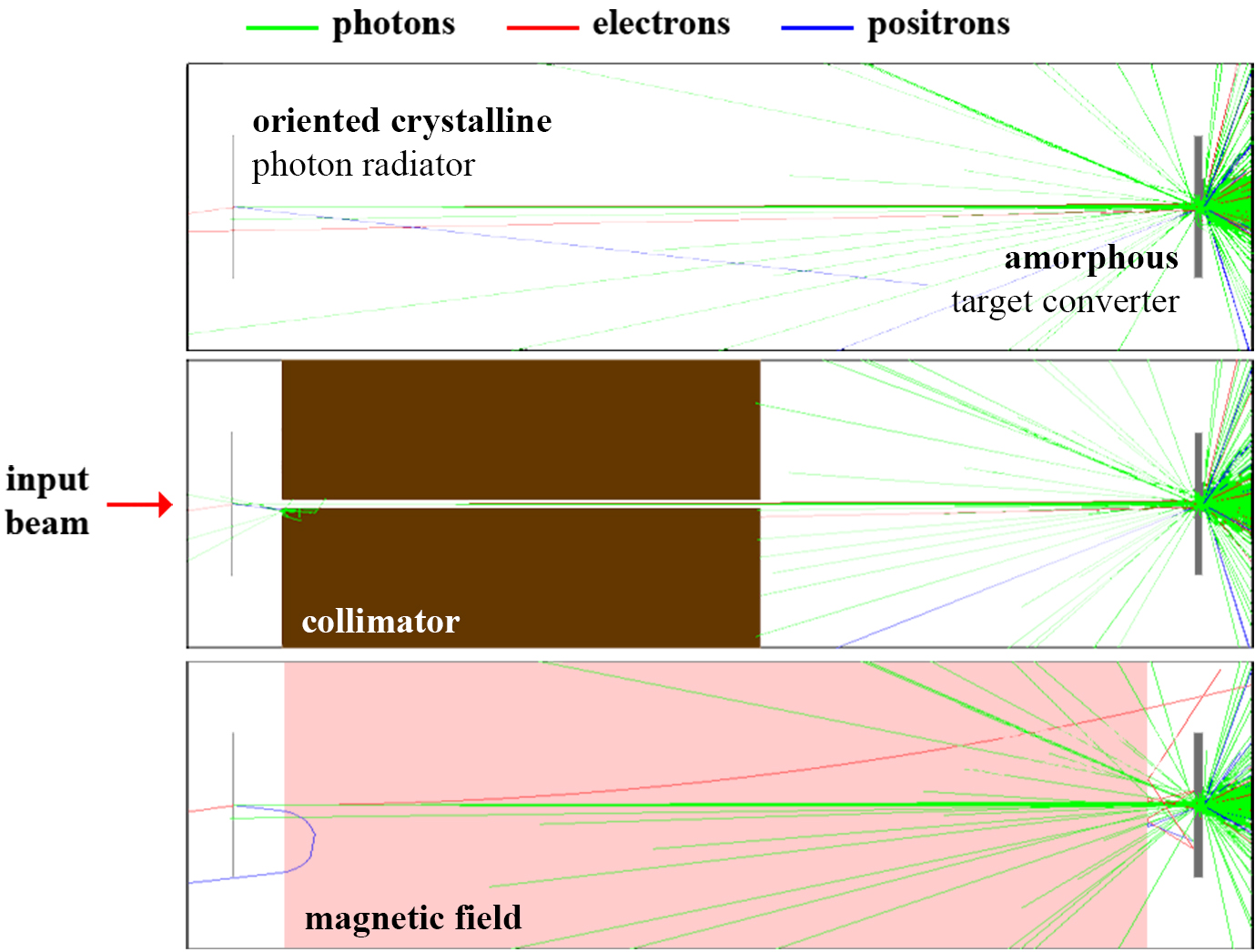}
\caption{Rendering of the target stage of the positron source Geant4 simulation, with 50 crystal output events displayed, in different schemes: hybrid simple (top) and hybrid optimised with collimators (centre) or with a bending magnet (bottom). The crystal bulk is not included in the simulation geometry and has been added here for the sake of visualisation. The longitudinal and transverse axes are not to scale.}
\label{fig:psFullSim_render_noMagn_all}
\end{figure}

Several different hybrid scheme configurations have been tested. The graphical representation of some of these configurations is presented in figure \ref{fig:psFullSim_render_noMagn_all}. The aim was to study how the positron source output performances vary as a function of the distance between the crystal and the amorphous target, $D$, of the thickness of the latter, $L$, and of the presence of a collimator (with variable square-shaped aperture $a$) or of a bending magnet. On the other hand, the shape and transverse size of the target have not been varied: all the runs have been performed on a parallelepiped-shaped target with a $199.75 \times 199.75~\mathrm{mm}^2$ square section. A very large target was purposely chosen, in order to properly study the energy distribution inside its volume up to several units of Molière radius from the beam centre.

Many different quantities have been scored: 

\begin{itemize}
\item the particle type, position and momentum of all the tracks at the converter rear face;

\item the integral energy deposited inside the whole target volume;

\item the integral energy deposited in each voxel -- parallelepiped-shaped, $\Delta x = \Delta y = 250~\mu\mathrm{m}$ (transverse) and ${\Delta z = 500~\mu\mathrm{m}}$ (longitudinal) -- of a mesh defined on top of the target volume.
\end{itemize}

\begin{figure}[h!]
\centering
\includegraphics[width=\columnwidth]{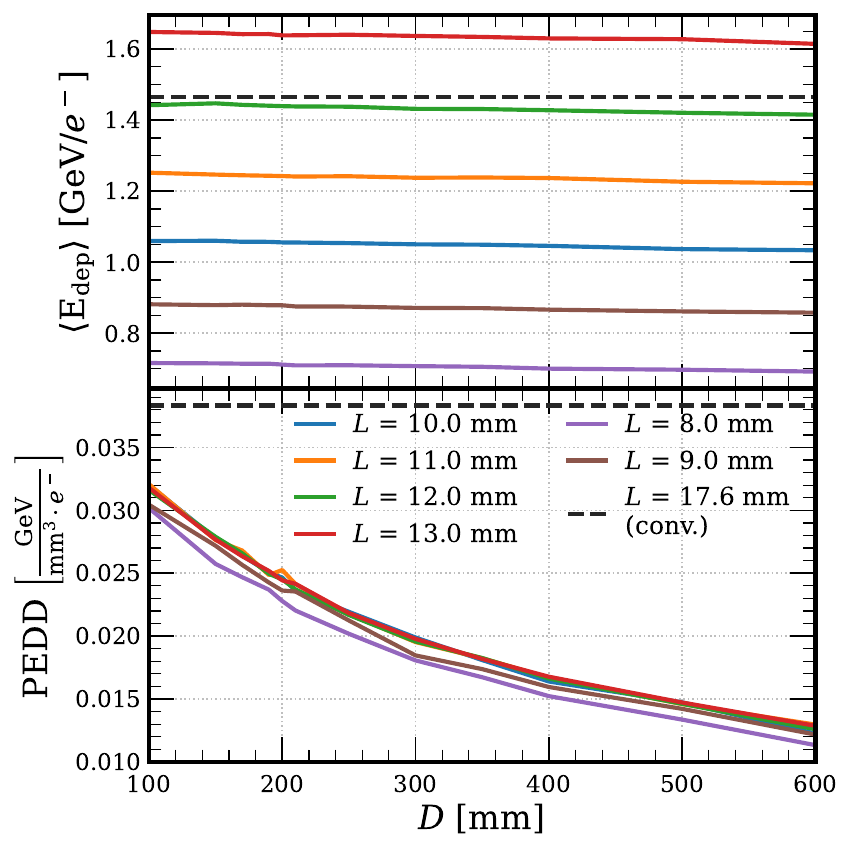}
\caption{Energy deposit (top) and PEDD (bottom) in the amorphous converter, as a function of $D$ and $L$. The corresponding values for the conventional positron source scheme are shown as black dashed lines.}
\label{fig:psFullSim_eDep_PEDD}
\end{figure}

Firstly, the no-collimator, no-magnet configuration will be discussed. Figure \ref{fig:psFullSim_eDep_PEDD} shows the results obtained from the target bulk as a function of $D$ and $L$. The plot at the top shows the energy deposit averaged on all the particle tracks. The value grows with the target thickness. On the other hand, all the curves depend on $D$ only slightly: a decrease of $\sim 2$--$3\%$ is observed as $D$ is increased from $10~\mathrm{cm}$ to $60~\mathrm{cm}$.

Moreover, the PEDD trends are shown in figure \ref{fig:psFullSim_eDep_PEDD} bottom. As the target is pulled away from the radiator, its angular acceptance decreases and a larger fraction of the crystal output beam tails is cut off. As a result, only the particles with the highest energy and hence the smallest angle (as shown in figure \ref{fig:psFullSim_crys_angles}) impinge on the target. On the other hand, a dependence on $L$ is only observed for values smaller than the depth along the beam direction at which the electromagnetic cascade reaches its peak in energy density per unit depth, which has been found at $\sim 9.5~\mathrm{mm}$ \cite{soldani_innovative_2023}. For every $L$ larger than this threshold value, the energy density is lower than the one at the threshold, therefore the PEDD corresponds to the energy density at $\sim 9.5~\mathrm{mm}$ independently on $L$; on the other hand, for smaller values of $L$, the PEDD always corresponds to the energy density attained towards the rear end of the bulk.

The energy deposit resulting from the simulation of the conventional scheme (with $L = 17.6~\mathrm{mm}$ -- black dashed line in the figures) is similar to the value obtained in the hybrid case with $L = 12~\mathrm{mm}$. This reflects the fact that, albeit thinner, the converter is hit by a large number of particles per primary electron, with an average energy much smaller than the one of the incident beam. On the other hand, all the hybrid-case PEDD values are considerably smaller than the one obtained in the conventional case.

\begin{figure}[h!]
\centering
\includegraphics[width=\columnwidth]{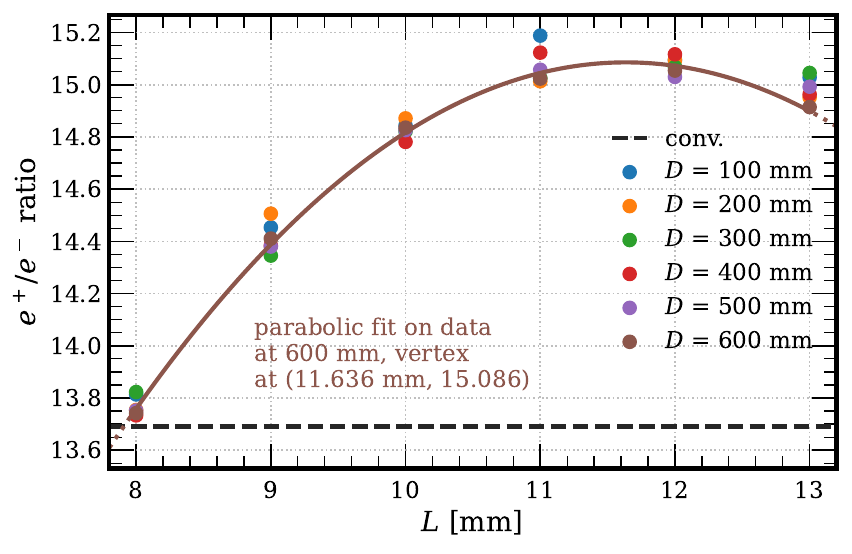}
\caption{Positron production rate from the hybrid positron source setup, as a function of $L$ and $D$. Only the positrons crossing the ($\sim 20 \times 20~\mathrm{cm}^2$) converter rear face are taken into account. The corresponding value for the conventional scheme is shown as a black dashed line.}
\label{fig:psFullSim_posRate}
\end{figure}

\begin{table*}[t!]
\centering
\boldmath
\begin{tabular}{r|cccccccc}
\textbf{Scheme}                                                              & \multicolumn{1}{c|}{conventional} & \multicolumn{7}{c}{hybrid}                        \\ \hline
\textbf{$L_\mathrm{crys}~\left[ \mathrm{mm} \right]$}                                           & \multicolumn{1}{c|}{--}  & \multicolumn{7}{c}{2}                           \\ \hline
\textbf{$D~\left[ \mathrm{m} \right]$}                                                  & \multicolumn{1}{c|}{--}  & \multicolumn{1}{c|}{0.6} & \multicolumn{3}{c|}{1} & \multicolumn{3}{c}{2} \\ \hline
\textbf{$L~\left[ \mathrm{mm} \right]$}                                                  & \multicolumn{1}{c|}{17.6} & \multicolumn{7}{c}{11.6}                         \\ \hline
\textbf{\begin{tabular}[c]{@{}r@{}}$\mathrm{Collimator}$ $(a = 5.5~\mathrm{mm})$\end{tabular}}                                                           & no             & no            & no   & yes  & no  & no  & yes  & no  \\ \hline
\textbf{\begin{tabular}[c]{@{}r@{}}$\mathrm{Magnet}$  $(B = 100~\mathrm{T} \times 90~\mathrm{cm})$\end{tabular}}                                                             & no             & no            & no   & no  & yes  & no  & no  & yes  \\ \hline
\textbf{$E_\mathrm{dep}$ in converter~$\left[ \mathrm{GeV} / e^- \right]$}                                        & 1.46            & 1.34           & 1.32  & 1.13 & 1.32 & 1.27 & 1.11 & 1.27 \\ \hline
\textbf{\begin{tabular}[c]{@{}r@{}}$\mathrm{PEDD}$ in converter~$\left[ \mathrm{MeV} / ( \mathrm{mm}^3 \cdot e^- ) \right]$\end{tabular}} & 38.3            & 12.8           & 8.4  & 8.2  & 8.4  & 4.1  & 3.8  & 3.9  \\ \hline
\textbf{Positron production rate}                                                         & 13.7            & 15.1           & 15.1  & 13.6 & 15  & 14.9 & 13.7 & 14.9 \\ \hline
\textbf{\begin{tabular}[c]{@{}r@{}}Output $e^+$ beam size (Gaussian sigma) $\left[ \mathrm{mm} \right]$\end{tabular}}                   & 0.7            & 1.0            & 1.2  & 1.2  & 1.2  & 1.5  & 1.5  & 1.5  \\ \hline
\textbf{\begin{tabular}[c]{@{}r@{}}Output $e^+$ beam divergence (Gaussian sigma) $\left[ \mathrm{mrad} \right]$\end{tabular}}                  & 25.9            & 27.4           & 26.8  & 27.7 & 28.9 & 29.2 & 25.6 & 27.1 \\ \hline
\textbf{\begin{tabular}[c]{@{}r@{}}Output $e^+$ mean energy $\left[ \mathrm{MeV} \right]$\end{tabular}}                  & 48.7            & 46.2           & 45.6  & 47.4 & 45.9 & 46.1 & 47.7 & 46.3 \\ \hline
\textbf{Neutron production rate}                                                          & 0.37            & 0.31           & 0.31  & 0.27 & 0.29 & 0.29 & 0.26 & 0.30  \\ \hline
\textbf{Photon production rate}                                                       & 299            & 310           & 308  & 270  & 307  & 301  & 268  & 301 
\end{tabular}
\unboldmath
\caption{Summary of the fully-optimised positron source full simulation results in different setup configurations.}
\label{tab:psFullSim_final}
\end{table*}

As mentioned above, the energy deposit and the PEDD must be kept at a reasonable level, while trying to achieve as high as possible a value of the positron yield per incident electron at the positron source output -- namely, the positron production rate, $e^+/e^-$. The estimates of the latter obtained counting the positrons at the ($\sim 20 \times 20~\mathrm{cm}^2$) converter rear face are shown in figure \ref{fig:psFullSim_posRate}.

Each dataset (obtained at a different $D$) shows a maximum in the positron production rate. The position of the maximum is independent on $D$ and corresponds to $L \sim 11.6~\mathrm{mm}$. Furthermore, the positron yield does not show any clear dependence on $D$. On the other hand, the yield grows with the output spatial acceptance: the comparison between the production rates estimated taking into account all the positrons crossing the converter rear plane in the whole simulation environment ($2.5 \times 2.5~\mathrm{m}^2$) and inside the converter rear face only (which is closer to the acceptance of a realistic magnetic capture system) shows a rate reduction of $\sim 5\%$ \cite{soldani_innovative_2023}.

The kinematic phase space at the system downstream end has also been studied \cite{soldani_innovative_2023}. The output positron divergence does not depend on $D$ and depends only weakly on $L$, and there is no substantial difference between the conventional and hybrid cases. On the other hand, the beam spot becomes larger as $D$ grows and also, less significantly, as $L$ grows. However, in all the simulated scenarios, the beam size does not increase significantly with respect to the input value ($500~\mu m$): a Gaussian sigma of $\sim 1~\mathrm{mm}$ is obtained at $60~\mathrm{cm}$, $\sim 1.2~\mathrm{mm}$ at $1~\mathrm{m}$ and $\sim 1.5~\mathrm{mm}$ at $2~\mathrm{m}$.

Both the mean (between $45$ and $50~\mathrm{MeV}$ in all the configurations under study) and the standard deviation of the positron energy spectra grow as a function of $L$ and are independent on $D$. Moreover, the standard deviation is in general approximately twice the corresponding mean value, which indicates the fact that the energy distributions at the positron source output are rather broad.

As shown in figure \ref{fig:psFullSim_render_noMagn_all} centre and bottom, simulations of the hybrid scheme with a collimator and with a magnetic field between the two targets have also been performed. The collimator consists of a $50~\mathrm{cm}$ thick block of tungsten, with transverse size ${2.5 \times 2.5~\mathrm{m}^2}$ and a square-shaped aperture along the beam axis; it is located with its front face at $5~\mathrm{cm}$ from the radiator rear face. The optimal value for the aperture size $a$ has been found at $5.5~\mathrm{mm}$, as it results in the same positron production rate as the conventional scheme with significant reduction of energy deposit and slight reduction of the PEDD with respect to the no-collimator case.

In case of the bending magnet, an ideal, $100~\mathrm{T}$ magnetic field directed along the vertical axis has been implemented, with the aim of sweeping away all the charged particles from the crystal regardless of their energy. The field is uniform and spans in a $90~\mathrm{cm}$ long region centered at $D/2$ from the crystal rear face.

\section{Results, conclusions and outlook}

Table \ref{tab:psFullSim_final} summarises all the main results obtained in the previous sections. All the variables of interest are provided for the most interesting positron source configurations that have been investigated. This table may be meant as an extension and an update of the results of \cite{bandiera_crystal-based_2022}.

In general, the crystal-based hybrid scheme options significantly improve the performance of the conventional scheme from the standpoint of all the features of interest in the positron source system design. In particular,

\begin{itemize}
\item increasing the distance between the crystalline radiator and the amorphous target strongly reduces the PEDD, whereas the positron production rate is slightly decreased, mostly due to leakage, and the positron beam size becomes slightly larger;

\item implementing a collimator results in a major reduction of the integral energy deposit;

\item implementing a magnetic field does not result in any significant improvement of the positron source performance except for a slight reduction of the PEDD (only observed with ${D = 2~\mathrm{m}}$);

\item no clear dependence on the configuration is observed in the positron output divergence and mean energy;

\item similarly, no major variation is observed in the numbers of neutrons and photons (of any energy) exiting from the target rear face per primary electron, which can serve as preliminary estimates of the amount of radiation in the environment surrounding the positron source.
\end{itemize}

Starting from the PEDD per incident electron reported in table \ref{tab:psFullSim_final} in all the selected configurations, it is possible to calculate the corresponding values per beam pulse for the lepton collider of interest. Considering the FCC-$ee$, it can be assumed that each pulse consists of two bunches of $2.1 \times 10^{10}$ electrons \cite{chaikovska_positron_2022}, which leads to a PEDD per pulse of $13.39~\mathrm{J/g}$ for the conventional scheme and of a range of values between $4.47~\mathrm{J/g}$ ($D = 60~\mathrm{cm}$, no collimator, no field) and $1.33~\mathrm{J/g}$ ($D = 2~\mathrm{m}$ with collimator) for all the investigated hybrid scheme options. All these values are significantly lower than the safety limit of $35~\mathrm{J/g}$ empirically found for tungsten \cite{chaikovska_positron_2022}; moreover, all the values obtained with the hybrid schemes are lower than the upper limit currently set for the first stage of the FCC-$ee$, $\sim 10.5~\mathrm{J/g}$ \cite{craievich_fcc-ee_nodate}.

All the simulation results presented in this work prove the potential of a hybrid positron source scheme that exploits an oriented crystalline radiator. Moreover, they can provide useful information for the design of the magnetic capture system that collects the positrons generated in the positron source into a low-emittance beam. The final choice of the parameters with which to develop the FCC-$ee$ positron source will heavily depend on the features of this capture system, which is currently at the design stage.

\section*{Acknowledgements}

We acknowledge support by the INFN CSN 1 and 5 (OREO, RD-FCC and RD-MUCOL projects), by the Ministero dell'Università e della Ricerca (PRIN 2022Y87K7X) and by the European Commission (Horizon 2020 AIDAinnova, GA 101004761; Horizon 2020 MSCA IF Global TRILLION, GA 101032975; Horizon EIC Pathfinder Open TECHNO-CLS, GA 101046458).

\bibliography{mybibfile}

\end{document}